\documentclass[pdflatex,sn-basic]{sn-jnl}

\usepackage{graphicx}%
\usepackage{multirow}%
\usepackage{amsmath,amssymb,amsfonts}%
\usepackage{amsthm}%
\usepackage{mathrsfs}%
\usepackage[title]{appendix}%
\usepackage{xcolor}%
\usepackage{textcomp}%
\usepackage{manyfoot}%
\usepackage{booktabs}%
\usepackage{algorithm}%
\usepackage{algorithmicx}%
\usepackage{algpseudocode}%
\usepackage{listings}%

\theoremstyle{thmstyleone}%
\theoremstyle{thmstyletwo}%

\theoremstyle{thmstylethree}%

\begin{document}

\title[Article Title]{Chemical Evolution of Galaxies: Past, Present and Future}


\author*[1,2,3,4]{\fnm{Francesca} \sur{Matteucci}}\email{francesca.matteucci@inaf.it}



\affil*[1]{\orgdiv{Department of Physics}, \orgname{Trieste University}, \orgaddress{\street{Via Fabio Valerio 2}, \city{Trieste}, \postcode{34143}, \state{Friuli-Venezia-Giulia}, \country{Italy}}}

\affil[2]{\orgdiv{Osservatorio Astronomico di Trieste}, \orgname{INAF}, \orgaddress{\street{Via G.B. Tiepolo, 11}, \city{Trieste}, \postcode{34131}, \state{Friuli-Venezia-Giulia}, \country{Italy}}}

\affil[3]{\orgdiv{Sezione di Trieste}, \orgname{INFN}, \orgaddress{\street{Via Fabio Valerio, 2}, \city{Trieste}, \postcode{34143}, \state{Friuli-Venezia-Giulia}, \country{Italy}}}

\affil[4]{\orgname{IFPU, Institute for the Fundamental Physics of the Universe}, \orgaddress{\street{Via Beirut, 2}, \city{Trieste}, \postcode{34151}, \state{Friuli-Venezia-Giulia}, \country{Italy}}}


\abstract{In this paper I will describe the basic principles of chemical evolution of galaxies, its main ingredients and  uncertainties. By means of chemical evolution we can perform the so-called galactic archaeology, which consists in reconstructing the history of star formation of galaxies and in particular of the Milky Way, starting from the observed stellar and gas abundances. Galactic archaeology is a powerful tool to predict also the behavior of galaxies at high redshift. In particular, we adopt the "time-delay model" which is a way of interpreting the [X/Fe] vs. [Fe/H] relations (X is the abundance of a specific chemical element) in terms of different timescales of the stellar progenitors of the chemical elements relative to Fe, which is the indicator of "metallicity". I will then describe how we can reconstruct the star formation histories of galaxies of different morphological type (spirals, ellipticals) starting from the available observations. Particular attention will be paid to the study of the Milky Way which is the best studied galaxy at the moment. I will start describing the first chemical evolution models and then the most recent ones, the main difference between the old and new models being the observational data to compare with. Finally, I will foresee which could be the future improvements to chemical models and what constraints can we derive to better understand galaxy evolution. }


\keywords{chemical evolution of galaxies, stellar nucleosynthesis, our Galaxy}



\maketitle

\section{Introduction}\label{sec1}

The chemical evolution of galaxies, namely the study of the evolution of gas and its chemical abundances, started with Beatrice Tinsley (\cite{tinsley1980} for a review), who posed the basis for the future development of this fascinating scientific field.  Galactic archaeology is based on the reconstruction of the history of formation of galaxies, and in particular our Galaxy, starting from the chemical abundances measured in stellar atmospheres and interstellar gas (HII regions and planetary nebulae).
Very few spectroscopic and photometric data were available at that time, if compared with the huge amount of data provided by modern surveys of the Milky Way: Gaia-ESO \citep{gilmore2012} APOGEE \citep{majewski2017}, 4MOST \citep{dejong2019}, MOONS \citep{cirasuolo2020}, LAMOST \citep{yu2021},   WEAVE \citep{jin2024}, MSE (MSE2019 Science Team), PFS \citep{chiba2026}, SDSS-V \citep{kollmeier2017}. However, in spite of that, the basic principles of chemical evolution were already established.
The papers \cite{tinsley1980}, \cite{greggio1983} and \cite{matteucci1986} all suggested the time-delay model, namely the way of interpreting the  observed [O/Fe] vs. [Fe/H] diagram for solar neighborhood stars, including members of the halo and disk. The first data were showing a plateau for the [O/Fe] ratio in stars with [Fe/H]$<$ -1.0 dex, namely in the halo stars, and a subsequent decline of that ratio for [Fe/H]$\ge$-1.0 dex, namely for the disk stars. The time-delay model interprets the observed behavior of the [O/Fe] ratio as due to the short lifetimes of the unique O producers (massive stars exploding as core-collapse supernovae), as opposed to the  long lifetimes of the main producers of Fe (Type Ia supernovae).
Type Ia supernovae (SNe), are believed to originate from white dwarfs in binary systems which explode in a time range from 30 Myr to a Hubble time, after the formation of the system. In this way, the [O/Fe] ratio in the early Galactic evolutionary phases, is dominated by the chemical enrichment from massive stars which are the main producers of O, and produce and eject only a small quantity of Fe. When Type Ia SNe start exploding, the Fe production increases and the [O/Fe] ratio decreases down to the solar value ([O/Fe]=0).  In the paper of \cite{matteucci1986}, for the first time the Type Ia SN enrichment was included in a detailed chemical evolution model, relaxing the Instantaneous Recycling Approximation (I.R.A.). In the following years, \cite{matteucci1990} predicted different behaviors for the [O/Fe] ratio in different environments, in particular in galaxies of different morphological type. They suggested that the "knee", where the slope of the [O/Fe] ratio changes, occurs at different values of [Fe/H] in objects with different star formation histories. In \cite{chiappini1997}, they tried to reproduce the data available at that time and built a model called "two-infall", where it was assumed that the stellar halo and thick disk formed out of a first infall of primordial gas occurring quite fast ($<$1 Gyr), followed by a second episode of primordial gas infall occurring on a much longer timescale ($\ge$7 Gyr in the solar vicinity) forming the thin disk. Starting from the years 2000's, high resolution data started to increase \citep{cayrel2004} as well as chemical evolution models trying to reproduce the observed abundance patterns of not only O but also other  $\alpha$-elements (C, Mg, Ne, Si, S, Ca), as well as Fe-peak elements \citep{francois2004}. In the following years (the 2010s),  large stellar spectroscopic Galactic surveys started with  Lamost (2012), Gaia-ESO (2013), GALAH(2015), APOGEE (2017). These large surveys have provided data for millions of stars and revealed features of the Galactic stellar populations ignored before, such as the observed bimodality of the [$\alpha$/Fe] ratios in the thick and thin disk stars. In fact, they have revealed two different [$\alpha$/Fe] sequences, almost parallel, identifying the thick disk stars as high-$\alpha$  and the thin disk ones as low-$\alpha$. Various papers tried to interpret this feature and among the suggested explanations we remind the existence of a gap in the star formation between thick and thin disk (see \cite{spitoni2019,spitoni2024}) and stellar migration from the innermost thin disk going to form the thick disk \citep{sharma2021}.\\
The evolution of the Galactic bulge was also studied in the past years starting with \cite{matteucci1990} and many following papers \citep{ballero2007,cescutti2011,grieco2012,matteucci2019}, among others. More data appeared later with \cite{hill2011} and \cite{rojas2017}. These data have shown that the stellar metallicity distribution function in the bulge shows two peaks, indicating either a stop in the star formation during the bulge assembly, or that stars were accreted by the bulge from the inner thin disk by means of the bar \citep{matteucci2019, molero2024}. It is important to note that all the chemical models for the bulge have suggested a very short timescale for its formation ($\sim$500 Myr).\\
Present and future stellar surveys  (WEAVE, MOONS, 4MOST, MSE, PFS, SDSS-V) will increase our knowledge about abundances in the Milky Way stellar populations.\\
Concerning other galaxies, we have high resolution  spectroscopic data for Local Group galaxies, either spirals  or irregulars and dwarf spheroidals, whereas for ellipticals we still rely on metallicity indicators derived from integrated spectra.\\
In this paper, we will show some highlights for the Milky Way and ellipticals, and demonstrate how the time-delay model interpretation is a robust tool to reconstruct the star formation history of galaxies as well as infer the morphology of high redshift galaxies.\\
The paper is organized as follows: in Section 2 we will describe the basic principles of chemical evolution models. In Section 3 the chemical evolution of the Milky Way, in Section 4 we will discuss elliptical galaxies, while in Section 5 we will present some future prospects.

\section{Chemical evolution models\label{sec2}}
Chemical evolution models need some basic ingredients, they are:
\begin{itemize}
\item Initial conditions: gas present since the beginning or accreted subsequently. Primordial gas or pre-enriched by Population III stars.
\item The history of star formation, namely the birthrate function:
\begin{equation}
B(m,t) =\psi(t) \varphi(m)dm dt,
\end{equation}

which depends upon the stellar mass and time. We use to separate the $B(m,t)$ into two functions, one only function of time, the star formation rate (SFR), $\psi(t)$, and one only function of stellar mass, the initial mass function (IMF), $\varphi(m)$. In order to derive the IMF, one should make assumptions on the SFR and vice versa. In reality, each function can in principle depend on both m and t. The most common parametrization of the SFR is a power law depending on the gas density (Kennicutt-Schmidt law):
\begin{equation}
    \psi(t)=\nu \sigma(t)_{gas}^k,
\end{equation}
  where k=1.5$\pm$0.15 \citep{kennicutt1998}.  
  The IMF is also a power law of the type:
  \begin{equation}
      \varphi(m) \propto m^{(1+x)},
  \end{equation}
where $x$ is known as the Salpeter index \citep{salpeter1955}. \\

The IMF has been derived only for solar neighborhood stars  \citep{miller1979, scalo1986, chabrier2002} and we do not know whether it is a universal function, although \cite{kroupa2001} suggested a universal IMF (UIMF) with changes of slope at 0.5 $M_{\odot}$ and 0.08$M_{\odot}$. On the other hand, later \cite{weidner2005}  presented the integrated  galaxial IMF (IGIMF) that vary among galaxies, depending on the different star formation histories and  galaxy masses.\\

\item Stellar yields: the nucleosynthesis occurring inside stars is a crucial ingredient in chemical evolution models. By stellar yields we intend the masses of the different chemical elements synthesized inside stars and ejected at their death.
Stars of different masses produce different chemical species: massive stars ($M> 8M_{\odot}$) are responsible for the formation of $\alpha$-elements, r-process elements, some Fe and light s-process elements. Low and intermediate mass stars (LIMS) ($0.8< M/M_{\odot}<8$) are responsible for the production of He, $^{14}N$, $^{12}C$, heavy s-process elements and most of the Fe through Type Ia SNe, which are believed to originate from white dwarfs (with LIMS progenitors) in binary systems. The most common progenitor models for Type Ia SNe are the single degenerate (SD) and double degenerate (DD) scenario \citep{matteucci2021}. Massive stars instead end their lives as core-collapse SNe (CC-SNe). 
Concerning r-process elements, they can in principle be formed by CC-SNe, although many modelers agree on the fact that only r-weak process can be produced \citep{travaglio2004}, whereas heavy r-process should arise from merging neutron stars \citep{thielemann2011}, as also shown by the gravitational event GW170817 \citep{pian2017}. 

\item Possible gas flows: infall, outflow and radial gas flows can alter the distribution of chemical abundances in galaxies and in particular along the thin disk of the Milky Way. Primordial or slightly enriched gas infall  can explain the formation of galactic disks, while gas outflows or winds are a natural consequence of stellar and AGN feedback in galaxies. Finally, radial gas flows are the natural consequence of gas infall for the preservation of angular momentum \citep{lacey1985}.\\
  The infall rate can be expressed by means of an exponential law:
  \begin{equation}
      I(t)=ae^{-t/\tau},
  \end{equation}
  where $a$ is a parameter fixed by reproducing the total surface mass density at the present time, and $\tau$ is the time scale for gas accretion.\\
  The wind rate can be expressed as proportional to the SFR:
  \begin{equation}
      W(t)=\lambda \cdot \psi(t),
  \end{equation}
  with $\lambda$ being the mass loading factor and a free parameter.
\end{itemize}
Chemical evolution models  include all of these ingredients and solve a number of integro-differential equations equal to the number of chemical elements, going from H to uranium and beyond.

\section{The Milky Way}\label{sec3}
We remind that in the MW we can distinguish four main stellar populations residing in the halo, bulge, thick and thin disk. Large recent spectroscopic surveys of the MW have provided very large data samples for each Galactic component.
\subsection{The past models}\label{subsec2}
In \cite{matteucci1986} for the first time it was presented a detailed chemical evolution model for the Milky Way including the chemical enrichment from Type Ia SNe. The Type Ia SN rate was computed as suggested by \cite{greggio1983} in the framework of the SD model for their progenitors. The nucleosynthesis prescriptions for Type Ia SNe were those from \cite{nomoto1986}. In \cite{matteucci1986} it was shown what had been already suggested by \cite{tinsley1980}, namely that to explain the observed behavior of the [$\alpha$/Fe] vs. [Fe/H] relation, one should invoke the delay in Fe production by Type Ia SNe with respect to the $\alpha$-elements, which are produced very fast by CC-SNe. This interpretation is known as {\it time-delay model}, and thanks to it we can derive the timescales of the formation of the various Galactic stellar components, by using the [$\alpha$/Fe] abundance ratios as cosmic clocks.
In Figure 1, we show the \cite{matteucci1986} models compared to the available data at that time. The models are indicated by lines and Model $a$, which reproduces the data, assumes that CC-SNe produce the bulk of O and part of Fe ($\sim30\%$) on very short timescales, while Type Ia SNe produce most of Fe ($\sim70\%$) and no O, on longer timescales. On the other hand, Model $b$ arbitrarily assumes that Fe is entirely produced by Type Ia SNe and clearly it does not fit the data. The conclusion was, in fact, that we need Fe from both SN Types but in different proportions.
It is worth noting that the time at which [Fe/H]=-1.0 dex, which marks roughly the highest metallicity of halo stars, is 1 Gyr. This time can therefore be interpreted as the timescale of the formation of the inner stellar halo.

\begin{figure}[h]
\centering
\includegraphics[width=0.7\textwidth]{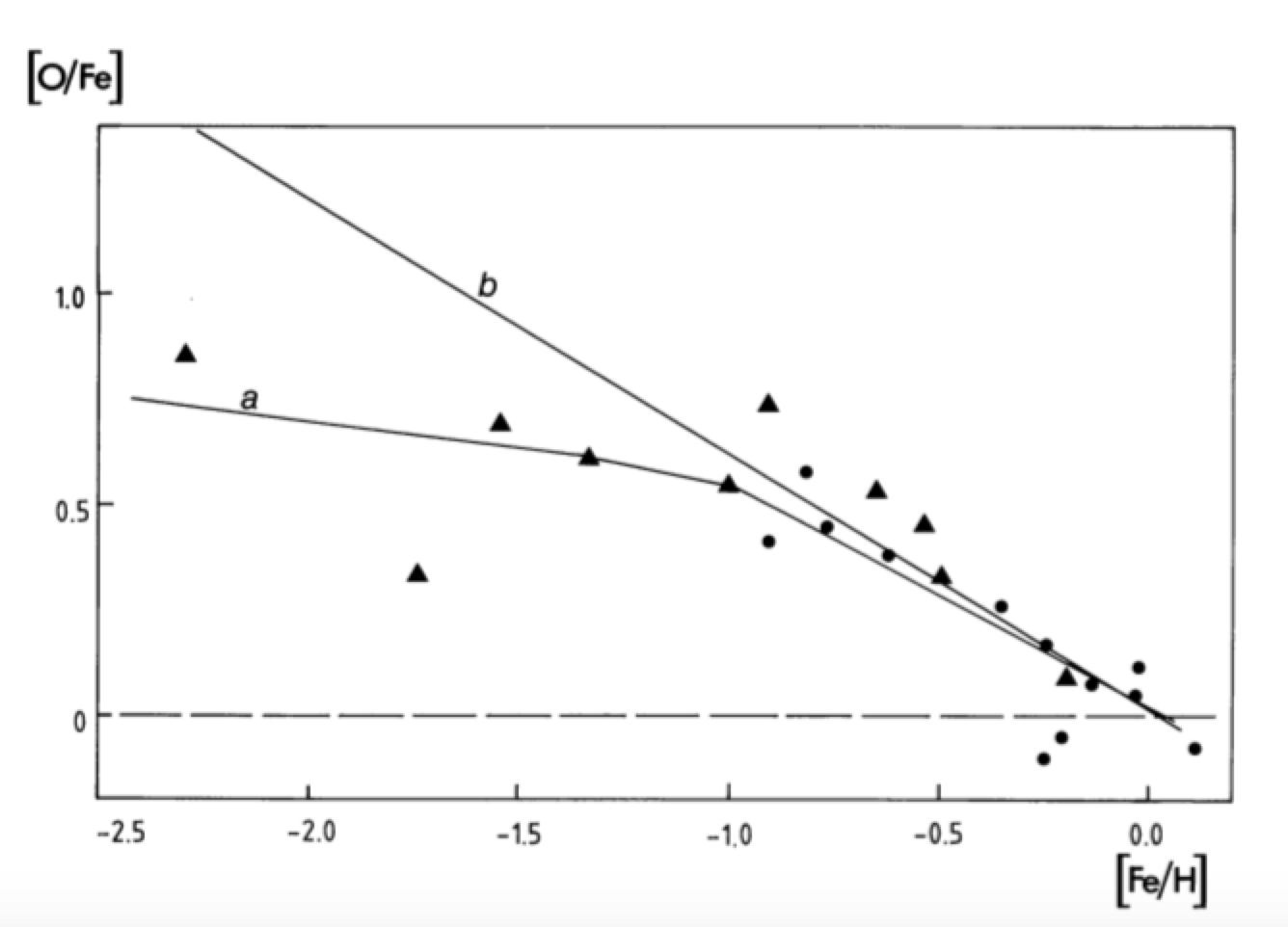}
\caption{The [O/Fe] vs. [Fe/H] diagram for the solar vicinity. The models and data are from \cite{matteucci1986}. The patterns shows a plateau for [Fe/H]$>$ -1.0 dex and a decline towards the solar value for larger metallicities. The interpretation of this behavior is the time-delay model: the plateau is due to CC-SNe which die immediately and form all the O and part of Fe, while the decline is due to the delayed contribution to Fe by Type Ia SNe which die with a time delay. $Model\, a$ is the best model whereas $Model\, b$ arbitrarily assumes that all of the Fe is produced by Type Ia SNe.}
\label{fig1}
\end{figure}

Later on, \cite{matteucci1990} described the [O/Fe] vs. [Fe/H] relation in different galactic environments such as spheroids/bulges, spirals  and irregular galaxies. In Figure 2, we show the predictions for the Galactic bulge, the solar vicinity and an irregular Magellanic galaxy. The upper curve refers to a spheroid like the Galactic bulge and predicts a long [O/Fe] plateau extending up to solar metallicity. This is due to the larger SFR, relative to the solar vicinity, assumed for such a system. This assumption was necessary to reproduce the observed metallicity distribution function (MDF) of bulge stars at that time \citep{rich1988}. The higher SFR leads, in fact, to a very fast increase of metallicity ([Fe/H]), so that when SNe Ia start to be important, the Fe abundance in the gas has already reached a high value thanks to the increased number of CC-SNe. The opposite occurs for SFRs milder than in the solar vicinity, as it happens for irregular galaxies and dwarf spheroidals: here the CC-SNe are less and when the Type Ia  SNe start exploding the gas metallicity is still quite low.  The plot of Figure 2 was a prediction, since at that time data for the bulge and Magellanic Clouds did not exist. \\
Later, it turned out that the predictions were correct, as shown in Figure 3, where some data are overimposed on the model predictions. It is worth noting that the curve referring to an irregular galaxy provides a good fit to data of LMC and Damped-Lymam $\alpha$- systems (DLAs), whose abundances of Fe should be corrected by dust depletion, as it is done in the data in Figure 3 \citep{vladilo2002}. These latter, because of their abundance patterns, are therefore irregular galaxies  \citep{matteucci1997}. Clearly, the [$\alpha$/Fe] vs. [Fe/H] plot can be used to infer the morphology of high redshift objects. Concerning the Galactic bulge, it is worth noting that more and more recent data than those shown in Figure 3 have confirmed the original predictions of \cite{matteucci1990}. In Figure 4 we show the results of \cite{cescutti2011} for the evolution of $\alpha$ elements and Fe in the bulge: the comparison between models and data confirmed the longer plateau expected for the [$\alpha$/Fe] ratios in the bulge. We remark that the best fit is provided by the Salpeter IMF, although for Si all the models predict an abundance higher than observed. This is likely due to the stellar yields of Si, which are still uncertain like those of many other species.

\begin{figure}
    \centering
    \includegraphics[width=0.7\linewidth]{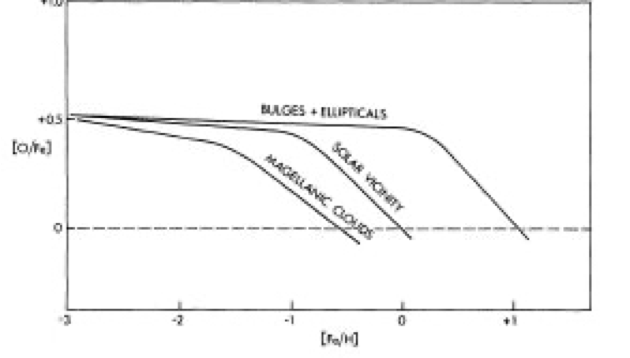}
     \caption{The [O/Fe] vs. [Fe/H] for three different galactic environments: the Galactic bulge, the solar vicinity and a Magellanic irregular galaxy. the upper curve refers to a spheroid like the Galactic bulge, where the SFR is very intense and the gas reaches quite soon the solar metallicity. On the other hand, the lowest curve refers to an irregular galaxy, where the SFR is milder than in the solar vicinity.
     Figure from \cite{matteucci1990}.}
    \label{fig2}
\end{figure}

\begin{figure}
        \centering
        \includegraphics[width=0.8\linewidth]{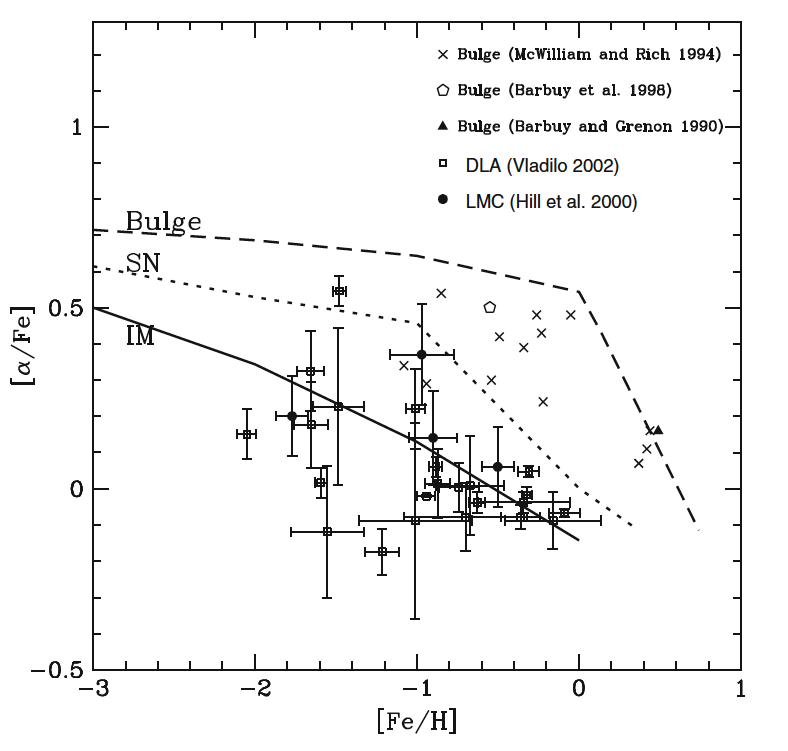}
        \caption{The predicted [$\alpha$/Fe] vs. [Fe/H] for three different galactic environments: the Galactic bulge, the solar vicinity and a Magellanic irregular galaxy, like in Figure 2, and compared to data. For $\alpha$ we intend a generic $\alpha$-element such as Mg or O. The data reported in the Figure are for the Galactic bulge \citep{mcwilliam1994,barbuy1998,barbuy1990} LMC \citep{hill2000} and DLAs \citep{vladilo2002}, as indicated in the figure.}
        
        \label{fig3}
    \end{figure}

All the models presented above are one-infall ones, namely they  assume that the MW formed as a result of one continuous infall episode and that the halo formed first, then the thick and finally the thin disk. The paper of \cite{chiappini1997} suggested instead the "two-infall model": in this model, the halo and thick disk were formed during a first fast gas infall episode, while the thin disk formed much more slowly as a result of a second independent infall episode. In between the two episodes, there was a period of negligible star formation. This model succeeded in reproducing most the properties of the various Galactic components, and this scenario has been adopted by many authors in the following years until recently.

\begin{figure}
    \centering
    \includegraphics[width=0.7\linewidth]{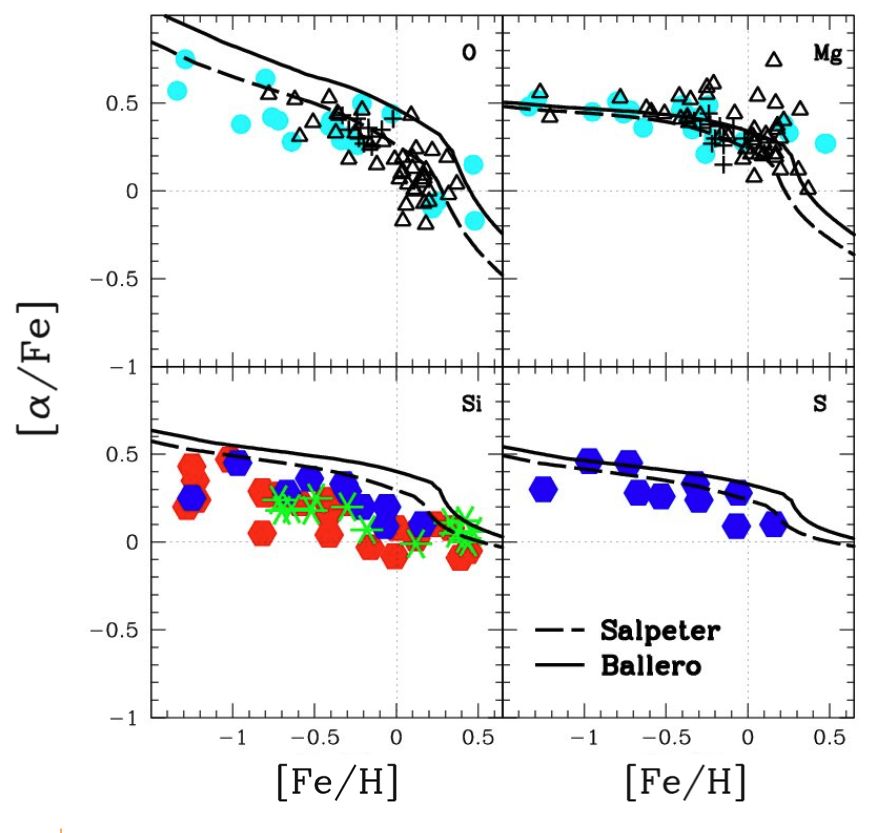}
    \caption{The [$\alpha$/Fe] vs [Fe/H] relation for the Galactic Bulge, compared with high resolution data. Figure from \cite{cescutti2011}. Here one can see that the "knee" in the relations occurs at metallicity higher than [Fe/H]=-1.0 dex, typical of the solar neighbourhood data. The two curves in each panel refer to different IMFs, both of which contain more massive stars than a typical IMF for the solar vicinity, as for example that of \cite{kroupa1993}.}
    \label{fig4}
\end{figure}

\subsubsection{The present models}\label{subsubsec2}

In the last years, several spectroscopic large surveys of the MW have provided data for millions of stars and allowed us to understand in more detail the formation and evolution of the  different Galactic components. Here, I will discuss the thick and thin disk as well as the bulge. The large surveys that I will use are APOGEE \citep{majewski2017}, Gaia-ESO \citep{gilmore2012} and AMBRE \citep{delaverny2013}.
These surveys are concentrated on the thick, thin disk and bulge stars. They have shown that the [$\alpha$/Fe] ratios in the thick and thin disk form two sequences and this has been called "bimodality of  the [$\alpha$/Fe] ratio". In Figures 5 and 6, we can see an example of that, with data taken from the AMBRE Survey. The two sequences for the thick and thin disk are evident. The stars of the thick disk (red points) are all $\alpha$-enhanced, while the stars of the thin disk (grey points) show lower [$\alpha$/Fe] ratios and the two sequences are almost parallel. This feature, if confirmed, tells us that the formation of the two disks cannot be explained by a simple sequential model. In Figures 5 and 6 are shown also some stars indicated with blue points that could belong to the thick disk or, in alternative,  be stars migrated from the inner Galactic disk.  In Figure 5, compared to the data are the predictions from the two-infall model, and it is evident the gap in the SFR between the thick and thin disk formation, appearing as a loop: in fact, when the SFR becomes negligible, at the end of the thick disk formation, the second gas infall dilutes the absolute abundances (in this case [Fe/H]), while it has no effect on the abundance ratios,  so that [Fe/H] decreases while [Mg/Fe] remains almost constant. Then, when the star formation resumes, the [Mg/Fe] ratio starts to decrease again due to the Fe produced from Type Ia SNe which did not stop exploding even during the star formation gap. In this case, the stars represented by the blue points can only be the results of migration from the innermost disk.
In Figure 6 we show the same data as in Figure 5 but compared with the predictions of the parallel model, where the two gas accretion events occur almost at the same time but at different rates. In this case, one can interpret the blue points in the plot as metal rich thick disk stars.

\begin{figure}[h]
    \centering
    \includegraphics[width=1.0\linewidth]{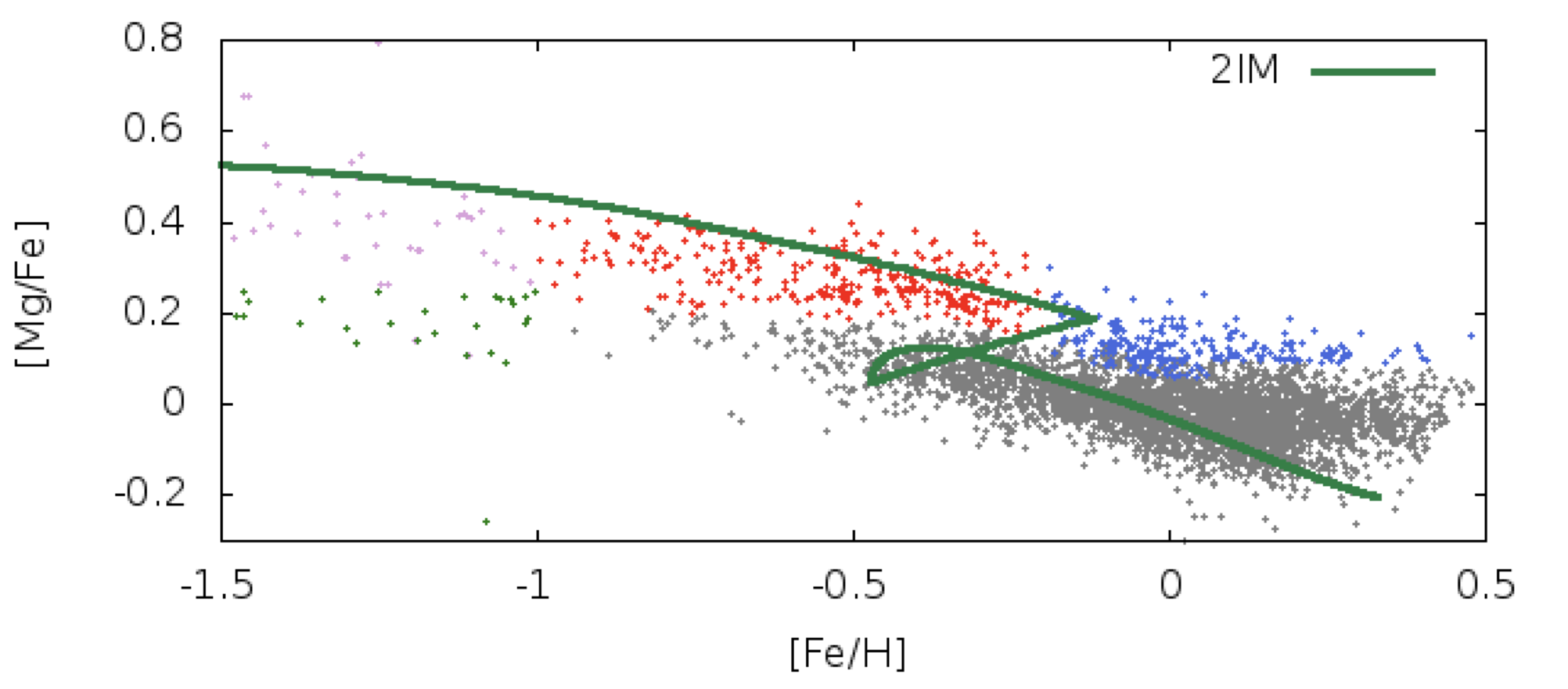}
    \caption{The observed [Mg/Fe] vs. [Fe/H] relation compared to the predictions of the two-infall model applied to the thick and thin disk. The data are from AMBRE Survey and the model from \cite{grisoni2017}.}
     \label{fig5}    
     \end{figure}    
     
     However, the pure parallel model has some limitations, since there should be at least a small time delay between the end for the formation of the thick and the beginning of the formation of the thin disk, otherwise it is difficult to understand from a physical point of view a long co-evolution period, where the two infall events should inevitably mix. On the other hand, recent papers \citep{borbolato2025} have suggested that the thin disk contains very old stars (age $>11$ Gyr). In any case, even in a situation of negligible star formation there would be a few stars forming anyway. Another warning about a pure parallel model, is represented by the fact that in this scenario it is difficult to explain the lower observed [$\alpha$/Fe] ratios at low [Fe/H] of thin disk stars, relative to thick disk ones at the same metallicity.

       \begin{figure}[h]
        \centering
        \includegraphics[width=1.0\linewidth]{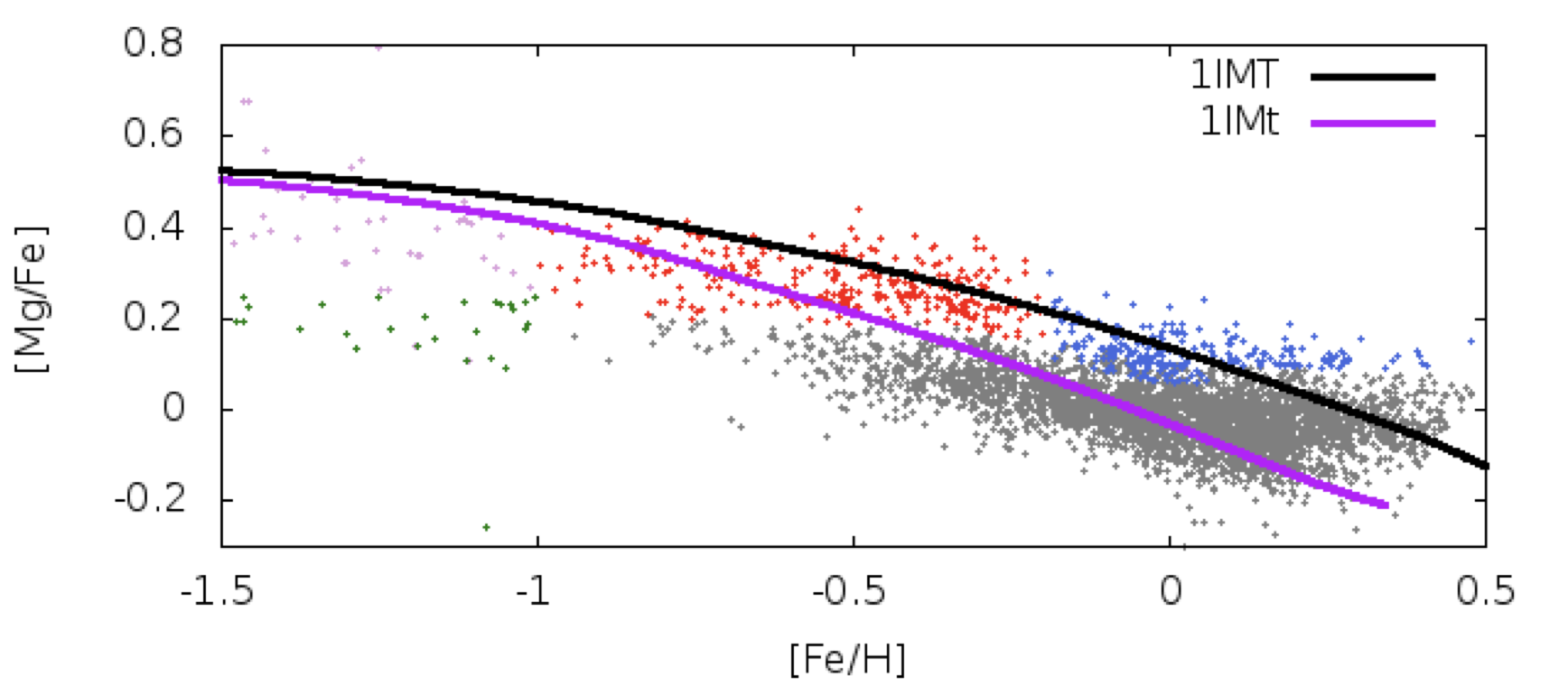}
        \caption{The observed [Mg/Fe] vs. [Fe/H] relation compared to the predictions of the parallel model applied to the thick and thin disk. The data are from AMBRE Survey and the model from \cite{grisoni2017}.}
        \label{fig6}
        \end{figure}

In 2019, \cite{spitoni2019} adopted the data of APOKASC \citep{silva2018}, based on APOGEE and Kepler, and interpreted the bimodality in the [Mg/Fe] ratio as due to a long gap in the star formation (roughly 4 Gyr) between the formation of the thick and thin disk. In \cite{spitoni2024}, the analysis was redone  with data DR17 of APOGEE \citep{abdurrouf2022} and showed that the gap is clearly visible in the APOGEE data, and in particular the stop in the star formation can be better appreciated if one plots [Fe/Mg] vs. [Mg/H] \citep{gratton1996,fuhrmann1998}.
In fact, in such a plot one can see an almost vertical increase of the [Fe/Mg] ratio at a fixed [Mg/H], indicating that the star formation has stopped and so the elements produced by CC-SNe  (such as O and Mg), while Fe continues to be produced by Type Ia SNe, thus making the ratio [Fe/Mg] to increase.  The model adopted in \cite{spitoni2024} is the two-infall one applied to the thick and thin disks with a gap in star formation of 3.5 Gyr, as derived by a Bayesian approach based on Mark of Chain Monte Carlo (MCMC) method. In Figure 7, we show the predictions of that model compared to the observations. In the same Figure, it is shown the stellar metallicity distribution function (MDF), as well the stellar distribution as a function of the [Fe/Mg] ratio. From the latter distribution it is clearly visible the bimodality (two peaks) in the [Fe/Mg] ratios.

        \begin{figure}
        \includegraphics[width=0.8\linewidth]{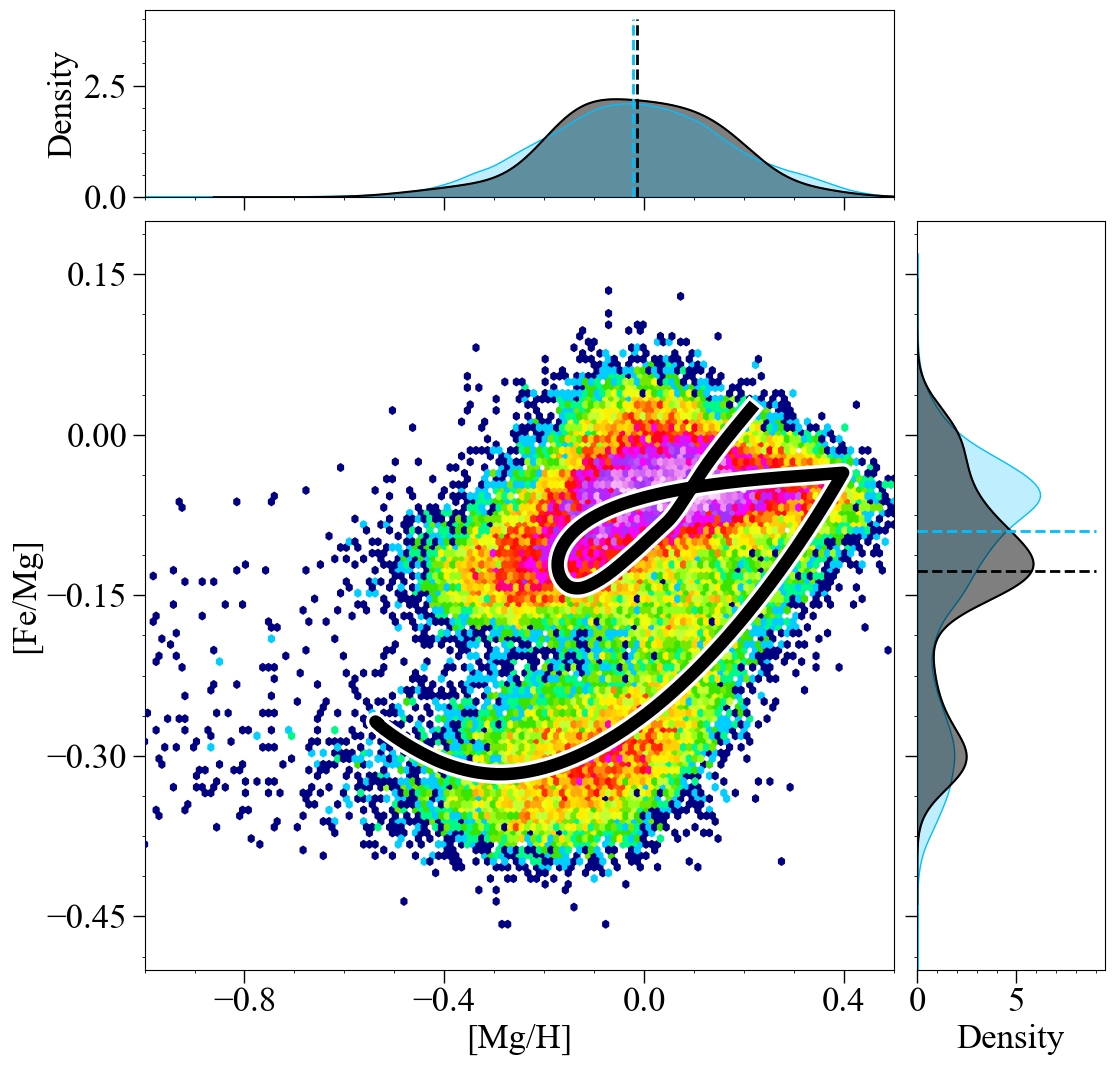}
        \caption{Predicted and observed [Fe/Mg] vs. [Mg/H] relation  for the two-infall model}. Data are from APOGEE DR17 and the model (continuous line) from \cite{spitoni2024}. Figure from \cite{spitoni2024}.
        \label{fig7}
         \end{figure}

Alternative explanations of the [$\alpha$/Fe] bimodality not considering a gap in the star formation nor multiple infall episodes are provided by \cite{schoenrich2009}, who invoked radial stellar mixing to explain the [$\alpha$/Fe] bimodality, or by \cite{sharma2021} who combined radial mixing with kinematic heating. Moreover, \citep{johnson2021, johnson2025} explained the bimodality as a result of a violent merger, in particular that of Gaia-Enceladus, while
\cite{hegedus2025} proposed a kind of two-infall model and \cite{dubay2026} questioned the two-infall model suggesting that it is in contrast with stellar ages. However, stellar ages still contain large uncertainties and other studies provide different results \citep{anders2023}.

\section{Elliptical galaxies}
\subsection{The old models}
   
In \cite{arimoto1986} (hereafter AY)  and \cite{matteucci1987} (hereafter MT) it was studied in detail the chemical evolution of elliptical galaxies, and in the latter paper, for the first time, the pollution from Type Ia SNe was included in models for ellipticals. The main conclusions of both papers were that elliptical galaxies evolved 
quite fast with a very rapid  metal enrichment and suffered galactic winds triggered by SN explosions: only by CC-SNe in  AY and by CC-SNe and Type Ia SNe in MT. After the galactic wind, the star formation was quenched and the galaxy evolved passively since then. In this scenario, ellipticals formed very fast and at very early epochs, at variance with the classical paradigm of the hierarchical galaxy formation scenario, where ellipticals, especially large ones, should have assembled recently by merging of smaller galaxies, as for example in \cite{thomas1999}. Recent observations by JWST have confirmed the existence of large galaxies at very high redshift \citep{curti2023}.

\begin{figure}
        \includegraphics[width=0.7\linewidth]{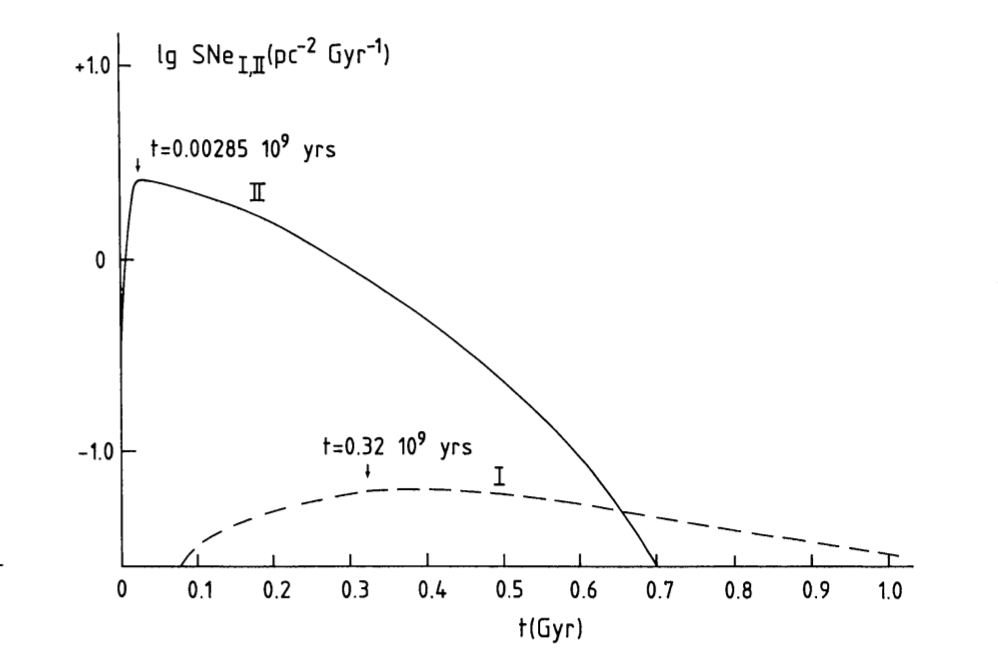}
        \caption{Predicted core-collapse and Type Ia SN rates in an elliptical galaxy with total stellar mass of $10^{12} M_{\odot}$. Figure from \cite{matteucci1987}. In the Figure are indicated the times at which each rate reaches its maximum.}
        \label{fig8}
         \end{figure}

On the other hand, Figure 8 shows that the Type Ia SN rate in ellipticals never stops, and these SNe continue to explode until the present time.
These continuous explosions inject large amounts of energy in the ISM and help to keep the galaxies quenched. In addition, Type Ia SNe are the main producers of the Fe observed in the intracluster medium  \citep{matteucci1988,mernier2016}.
In particular, in Figure 8 we show the predictions of MT for the CC- and Type Ia SN rates in a large elliptical galaxy with a stellar mass of $\sim 10^{12}M_{\odot}$. The adopted IMF is that of AY and is a top-heavy one (Salpeter index x=0.95). As one can see, the CC-SN rate stops when star formation dies, and this occurs when a galactic wind develops.
It is worth reminding that in ellipticals the abundances are measured only from the observed integrated spectra and that represent the dominant stellar population in the visual light. Therefore, they are not abundances of single stars, as in our Galaxy, and we refer to them as abundance indicators \citep{burstein1997}. Then, adopting appropriate calibrations, the indicators can be transformed into real abundances.
In \cite{matteucci1994}, it was demonstrated that, in order to reproduce the observed increase of the [$\alpha$/Fe] ratio with galactic mass in ellipticals, one should assume a down-sizing in star formation, in other words, an inverse wind scenario, where the most massive systems develop galactic winds and thus star formation stops, before less massive ones, in spite of the deeper potential well. This can be obtained by assuming that the efficiency of star formation is an increasing function of the galactic mass. In Figure 9, we show the plot [$\alpha$/Fe] vs. $\sigma$, with $\sigma$ being the velocity dispersion which is a tracer of the total stellar mass $M_*$, where  models with downsizing in star formation \citep{pipino2004} are compared to observational data and do reproduce the increase of the [$\alpha$/Fe] ratio in the more massive galaxies. This is due to the fact that the SFR stops earlier in more massive systems, and therefore the contribution to Fe from Type Ia SNe into the gas going to form stars, is less than in galaxies with a longer period of star formation. In Figure 9, the predictions of a typical hierarchical model indicating the opposite of what observed are also shown, and in particular decreasing ratios with increasing galactic mass; this is due to the assumed longer period of formation for the largest ellipticals relative to the smaller ones. Another possible explanation, also suggested in \cite{matteucci1994} is to assume a variation in the IMF among ellipticals of different mass, having more massive stars in more massive galaxies \citep{recchi2009,fontanot2017}.

       \begin{figure}[h]
        \centering
        \includegraphics[width=0.9\linewidth]{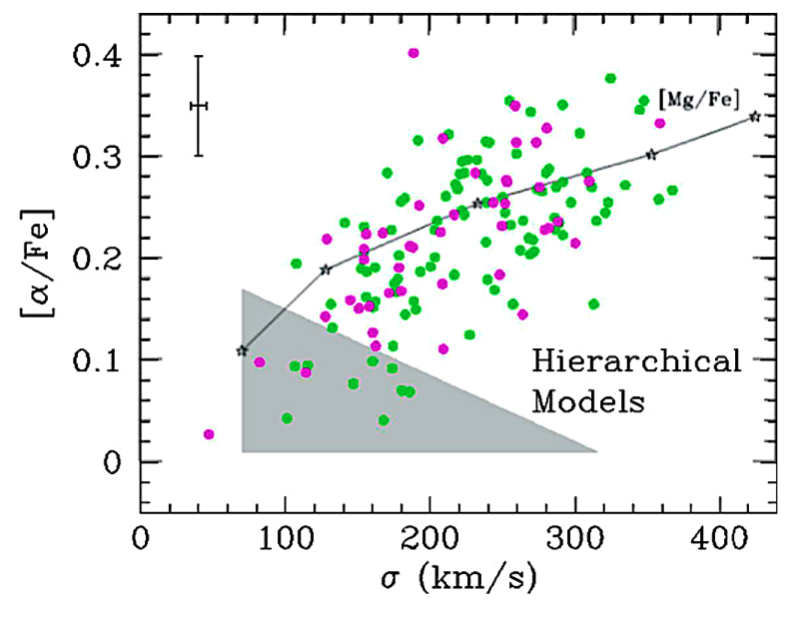}
        \caption{The [$\alpha$/Fe] vs. $\sigma$ (velocity dispersion) relation, predicted and observed for elliptical galaxies. The curve refers to the  predictions of the model of \cite{pipino2004}, whereas the grey triangle indicates the predictions of a typical hierarchical model of galaxy formation by \cite{thomas1999}. Figure adapted from \cite{thomas2002} and \cite{thomas1999}.}
        \label{fig9}
\end{figure}

In the following years, the downsizing in star formation for ellipticals was better established and some hierarchical models included this effect in modeling the evolution of these galaxies \citep{delucia2006, calura2011}.

\subsection{Present Models}
More recently, \cite{demasi2018,demasi2019} developed more refined models of chemical evolution of ellipticals aimed to reproduce the main chemical features of these galaxies, such as the mass-metallicity relation, mass-[$\alpha$/Fe] relation and radial abundance gradients.
In those papers, the effects of different IMFs were taken into account. In \cite{demasi2018}, it was suggested that to reproduce the features of ellipticals the IMF should be bottom-heavy in less massive galaxies and top-heavy in more massive ones. This is at variance with other suggestions \citep{ferreras2013} claiming a bottom- heavy IMF in massive ellipticals: \cite{demasi2018} showed that such IMF could not reproduce the observed increase of [$\alpha$/Fe] vs. $M_*$ (the galactic stellar mass). In the same paper, they also tested the integrated galaxial IMG (IGIMF) \citep{weidner2005} which depends on the SFR and can also reproduce the trend of [$\alpha$/Fe] , since it predicts a more top-heavy IMF in larger objects.\\
In Figure 10 we show an example of the observed  mass-metallicity relation in ellipticals, compared to the models of \cite{demasi2018}.

\begin{figure}[h]
        \centering
        \includegraphics[width=0.9\linewidth]{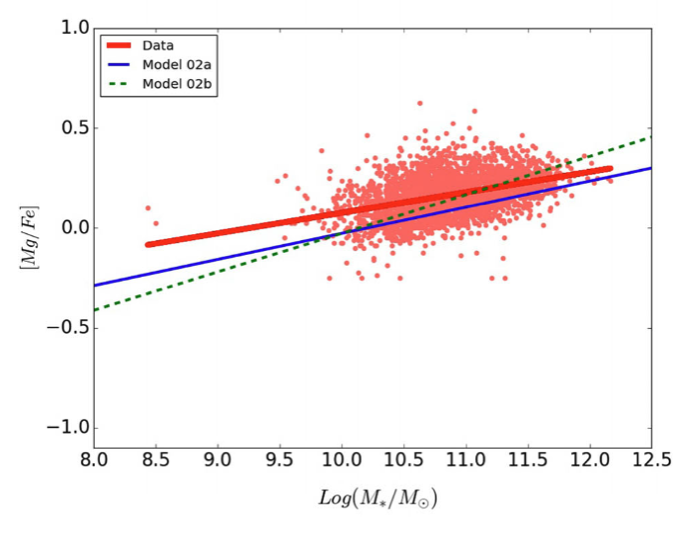}
        \caption{The [Mg/Fe] vs. $log(M_*/M_{\odot})$ relation for a sample of elliptical galaxies (red points). The best fit to the data is the thick red line, while the dotted  and light continuous ones represent predictions of models with variable IMF as  functions of $M_*$. 
        In particular, the dotted green line represents models adopting the IGIMF, while the continuous blue line is a combination of IMFs increasingly top-heavy from the less to the more massive galaxies. Data are from \cite{thomas2010}.
        Figure from \cite{demasi2018}.}
        \label{fig10}
\end{figure}

In \cite{molero2023}, a new chemical model for ellipticals was presented, including both stellar and AGN feedback. In the stellar feedback, supernovae and stellar winds were considered. They pointed out the fundamental role that Type Ia SNe play in keeping elliptical galaxies quenched for most of their life, and that AGN feedback is negligible in less massive galaxies, while it is important in the most massive ones to trigger the development of galactic winds. In these models, the efficiency of star formation increases with stellar mass, as in \cite{matteucci1994}, so to produce an inverse wind scenario (see Figure 9).  They ran models for galaxies with masses in the range $5\cdot10^{9}- 10^{12}M_{\odot}$, and were able to reproduce the increase of the [O/Fe] ratio with galactic stellar mass and also the Magorrian relation ($M_{BH}$ vs. $M_* (\sigma)$), as shown in Figure 11.


 \begin{figure*}[h]
 \centering
 \includegraphics[width=0.45\linewidth]{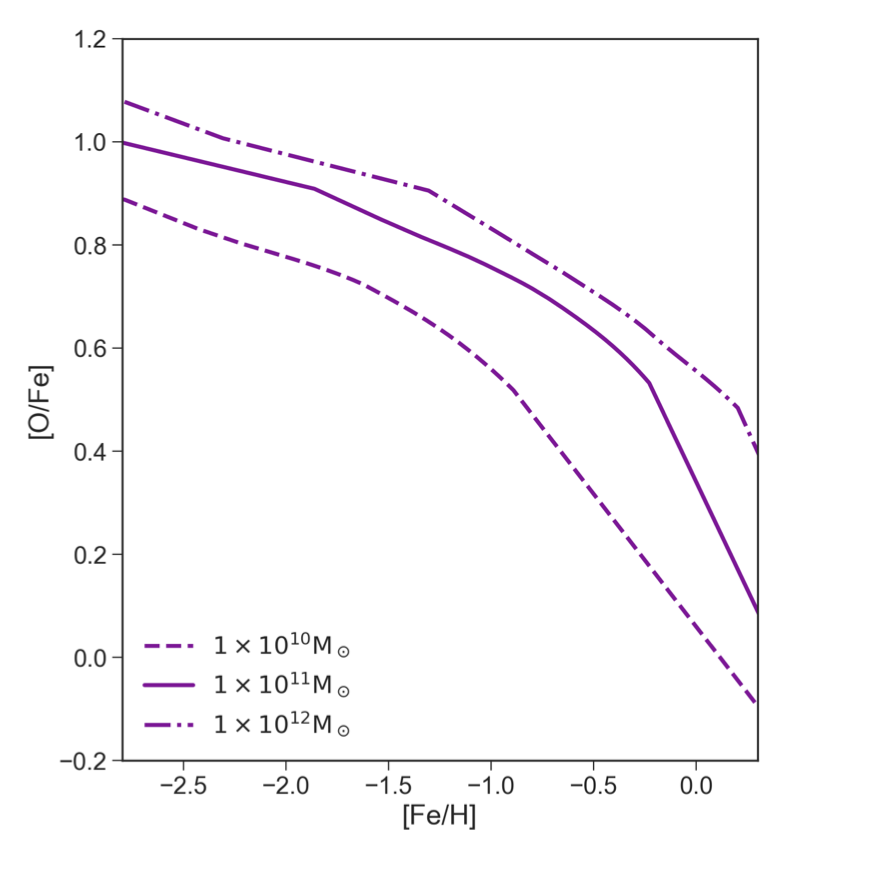}
 \includegraphics[width=0.45\linewidth]{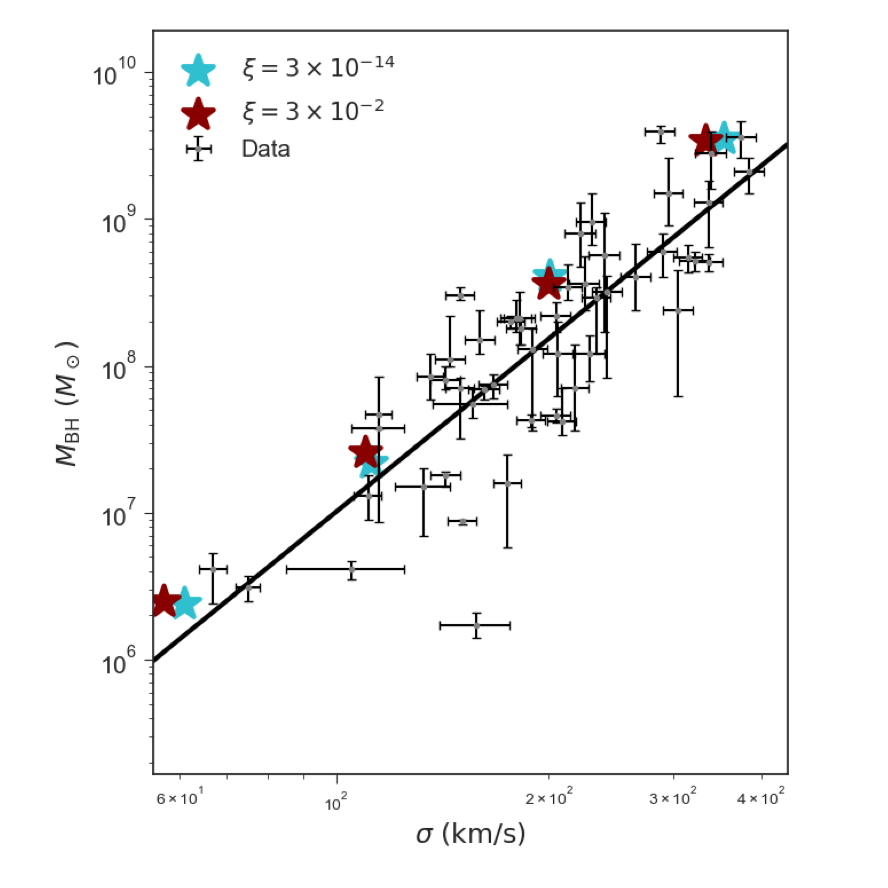} 
 \caption{Left panel: predicted behavior of the [O/Fe] ratio in the gas of ellipticals of different galactic stellar masses, as indicated in the figure. The star formation stops first in massive ellipticals, so clearly the [O/Fe] ratio in the stellar population dominating the visual spectrum is higher in more massive ellipticals, as observed.
 Right panel: predicted behavior of the $M_{BH}$ vs. $M_*$ compared to the data \citep{gultekin2010}. The grey points with error bars are the data, while the cyan and red stars represents the model predictions obtained for two different values of the absorption coefficient $\xi$ \citep{ciotti2001}, as indicated in the figure.  
 Figures from \cite{molero2023}.}
\label{fig11}
\end{figure*}


\section{The future}\label{sec12}

Galactic chemical evolution will certainly be one of the most important scientific subjects in the future, since more and more high resolution chemical abundances will be acquired by means of the largest optical and infrared telescopes. The advent of ELT will open unprecedented opportunities for studying the abundances in external galaxies. In this way, we will obtain better constraints from the chemical models, which will be more and more detailed in their inputs. Dynamical and chemical evolution will be coupled  to better understand galaxy formation. At the same time, stellar nucleosynthesis models will be more and more precise and will overcome the uncertainties still present in the calculation of stellar yields, which is one of the main parameters in galactic chemical evolution models. In fact, the nucleosynthesis of some $\alpha$- and Fe-peak elements, as well as that of some s- and r-process elements, still contains many uncertainties.\\
To conclude, I would like to highlight that, in spite of the rather simple approach followed by galactic chemical evolution models, they had and continue to have a good predictive power.\\

\section{Acknowledgements}
F. Matteucci  thanks I.N.A.F. for the 1.05.24.07.02 Mini Grant
- LEGARE "Linking the chemical Evolution of Galactic discs AcRoss diversE scales: from the thin disc to the nuclear stellar disc" (PI E. Spitoni).
F. Matteucci  also  thanks I.N.A.F. for the 1.05.12.06.05 Theory Grant - Galactic archaeology with radioactive and stable nuclei. 
F. Matteucci  also thanks support from Project PRIN
MIUR 2022 (code 2022ARWP9C) "Early Formation and Evolution of Bulge and HalO (EFEBHO)" (PI: M. Marconi). F. Matteucci thanks an anonymous referee for careful reading of the manuscript and useful suggestions.

\section*{Declarations}


\begin{itemize}
\item Funding

Not Applicable
\item Ethics approval and consent to participate

Not Applicable
\end{itemize}

\bibliography{sn-bibliography}

\end{document}